\documentclass[preprint2]{emulateapj}
\usepackage{color}
\usepackage{amsmath}
\usepackage{threeparttable}
\usepackage{natbib}
\usepackage{times}

\DeclareGraphicsExtensions{.jpg,.pdf,.png,.eps,.ps}

\newcommand{\nhat}{\hat{\mathbf{n}}}

\newcommand{\kvec}{\mathbf{k}}

\newcommand{\degs}{deg$^2$}

\def\oz#1{{}}

\def\ell{l}

\newcommand{\chicmb}{\chi_{\textrm{\tiny CMB}}}

\newcommand{\comment}[1]{{}}

\def\KICPChicago{1}
\def\PhysicsUChicago{2}
\def\McGill{3}
\def\UChicago{4}
\def\NCSA{5}
\def\CfA{6}
\def\EFIChicago{7}
\def\Michigan{8}
\def\Miss{9}
\def\Zurich{10}
\def\LBNL{11}
\def\AAUChicago{12}
\def\Argonne{13}
\def\NIST{14}
\def\Munich{15}
\def\ExcellenceCluster{16}
\def\Caltech{17}
\def\JPL{18}
\def\Berkeley{19}
\def\UFlorida{20}
\def\Colorado{21}
\def\SantaCruz{22}
\def\Davis{23}
\def\NASA{24}
\def\Arizona{25}
\def\MPE{26}
\def\CaseWestern{27}
\def\Minnesota{28}
\def\STScI{29}
\def\ArtInstChicago{30}
\def\Yale{31}
\def\LosAlamos{32}
\def\KIPAC{33}
\def\Stanford{34}
\def\SLAC{35}
\def\BCCP{36}

\begin{document}

\title{A Measurement of the Correlation of Galaxy Surveys with CMB Lensing Convergence Maps from the South Pole Telescope}

\slugcomment{}

\author{
 L.~E.~Bleem,\altaffilmark{\KICPChicago,\PhysicsUChicago}
 A.~van~Engelen,\altaffilmark{\McGill}
 G.~P.~Holder,\altaffilmark{\McGill}
 K.~A.~Aird,\altaffilmark{\UChicago}
 R.~Armstrong,\altaffilmark{\NCSA}
 M.~L.~N.~Ashby,\altaffilmark{\CfA}
 M.~R.~Becker,\altaffilmark{\KICPChicago,\PhysicsUChicago}
 B.~A.~Benson,\altaffilmark{\KICPChicago,\EFIChicago}
 T.~Biesiadzinski,\altaffilmark{\Michigan}
 M.~Brodwin,\altaffilmark{\Miss}
 M.~T.~Busha,\altaffilmark{\Zurich,\LBNL}
 J.~E.~Carlstrom,\altaffilmark{\KICPChicago,\PhysicsUChicago,\EFIChicago,\AAUChicago,\Argonne}
 C.~L.~Chang,\altaffilmark{\KICPChicago,\EFIChicago,\Argonne}
 H.~M. Cho,\altaffilmark{\NIST} 
 T.~M.~Crawford,\altaffilmark{\KICPChicago,\AAUChicago}
 A.~T.~Crites,\altaffilmark{\KICPChicago,\AAUChicago}
 T.~de~Haan,\altaffilmark{\McGill}
 S.~Desai,\altaffilmark{\Munich,\ExcellenceCluster}
 M.~A.~Dobbs,\altaffilmark{\McGill}
 O.~Dor\'e,\altaffilmark{\Caltech,\JPL}
 J.~Dudley,\altaffilmark{\McGill}
 J.~E.~Geach,\altaffilmark{\McGill}
 E.~M.~George,\altaffilmark{\Berkeley}
 M.~D.~Gladders,\altaffilmark{\KICPChicago,\AAUChicago}
 A.~H.~Gonzalez,\altaffilmark{\UFlorida}
 N.~W.~Halverson,\altaffilmark{\Colorado}
 N.~Harrington,\altaffilmark{\Berkeley}
 F.~W.~High,\altaffilmark{\KICPChicago,\AAUChicago}
 B.~P.~Holden,\altaffilmark{\SantaCruz}
 W.~L.~Holzapfel,\altaffilmark{\Berkeley}
 S.~Hoover,\altaffilmark{\KICPChicago,\PhysicsUChicago}
 J.~D.~Hrubes,\altaffilmark{\UChicago}
 M.~Joy,\altaffilmark{\NASA}
 R.~Keisler,\altaffilmark{\KICPChicago,\PhysicsUChicago}
 L.~Knox,\altaffilmark{\Davis}
 A.~T.~Lee,\altaffilmark{\Berkeley,\LBNL}
 E.~M.~Leitch,\altaffilmark{\KICPChicago,\AAUChicago}
 M.~Lueker,\altaffilmark{\Berkeley,\Caltech}
 D.~Luong-Van,\altaffilmark{\UChicago}
 D.~P.~Marrone,\altaffilmark{\Arizona}
 J.~Martinez-Manso,\altaffilmark{\UFlorida}
 J.~J.~McMahon,\altaffilmark{\Michigan}
 J.~Mehl,\altaffilmark{\KICPChicago}
 S.~S.~Meyer,\altaffilmark{\KICPChicago,\EFIChicago,\PhysicsUChicago,\AAUChicago}
 J.~J.~Mohr,\altaffilmark{\Munich, \ExcellenceCluster, \MPE}
 T.~E.~Montroy,\altaffilmark{\CaseWestern}
 T.~Natoli,\altaffilmark{\KICPChicago,\PhysicsUChicago}
 S.~Padin,\altaffilmark{\KICPChicago,\AAUChicago,\Caltech}
 T.~Plagge,\altaffilmark{\KICPChicago,\AAUChicago}
 C.~Pryke,\altaffilmark{\Minnesota }
 C.~L.~Reichardt,\altaffilmark{\Berkeley}
 A.~Rest,\altaffilmark{\STScI}
 J.~E.~Ruhl,\altaffilmark{\CaseWestern}
 B.~R.~Saliwanchik,\altaffilmark{\CaseWestern}
 J.~T.~Sayre,\altaffilmark{\CaseWestern}
 K.~K.~Schaffer,\altaffilmark{\KICPChicago,\EFIChicago,\ArtInstChicago}
 L.~Shaw,\altaffilmark{\Yale} 
 E.~Shirokoff,\altaffilmark{\Berkeley} 
 H.~G.~Spieler,\altaffilmark{\LBNL}
 B.~Stalder,\altaffilmark{\CfA}
 S.~A.~Stanford,\altaffilmark{\Davis}
 Z.~Staniszewski,\altaffilmark{\CaseWestern}
 A.~A.~Stark,\altaffilmark{\CfA}
 D.~Stern,\altaffilmark{\JPL}
 K.~Story,\altaffilmark{\KICPChicago,\PhysicsUChicago}
 A.~Vallinotto,\altaffilmark{\LosAlamos}
 K.~Vanderlinde,\altaffilmark{\McGill}
 J.~D.~Vieira,\altaffilmark{\KICPChicago,\PhysicsUChicago,\Caltech} 
 R.~H.~Wechsler,\altaffilmark{\KIPAC,\Stanford,\SLAC}
 R.~Williamson,\altaffilmark{\KICPChicago,\AAUChicago} and
 O.~Zahn\altaffilmark{\BCCP} 
}

\altaffiltext{\KICPChicago}{Kavli Institute for Cosmological Physics,
University of Chicago, 5640 South Ellis Avenue, Chicago, IL, USA 60637}

\altaffiltext{\PhysicsUChicago}{Department of Physics,
University of Chicago,
5640 South Ellis Avenue, Chicago, IL, USA 60637}

\altaffiltext{\McGill}{Department of Physics,
McGill University, 3600 Rue University, 
Montreal, Quebec H3A 2T8, Canada}

\altaffiltext{\UChicago}{University of Chicago,
5640 South Ellis Avenue, Chicago, IL, USA 60637}

\altaffiltext{\NCSA}{National Center for Supercomputing Applications,
University of Illinois, 1205 West Clark Street, Urbana, IL 61801}

\altaffiltext{\CfA}{Harvard-Smithsonian Center for Astrophysics,
60 Garden Street, Cambridge, MA, USA 02138}

\altaffiltext{\EFIChicago}{Enrico Fermi Institute,
University of Chicago,
5640 South Ellis Avenue, Chicago, IL, USA 60637}

\altaffiltext{\Michigan}{Department of Physics, University of Michigan, 450 Church Street, Ann  Arbor, MI, USA 48109}

\altaffiltext{\Miss}{Department of Physics and Astronomy, University of Missouri, 5110 Rockhill Road, Kansas City, MO 64110}

\altaffiltext{\Zurich}{Institute for Theoretical Physics, University of Z\"{u}rich, Z\"{u}rich, Switzerland}

\altaffiltext{\LBNL}{Physics Division,Lawrence Berkeley National Laboratory,Berkeley, CA, USA 94720}

\altaffiltext{\AAUChicago}{Department of Astronomy and Astrophysics,
University of Chicago,
5640 South Ellis Avenue, Chicago, IL, USA 60637}

\altaffiltext{\Argonne}{Argonne National Laboratory, 9700 S. Cass Avenue, Argonne, IL, USA 60439}

\altaffiltext{\NIST}{NIST Quantum Devices Group, 325 Broadway Mailcode 817.03, Boulder, CO, USA 80305}

\altaffiltext{\Munich}{Department of Physics, Ludwig-Maximilians-Universit\"{a}t,Scheinerstr.\ 1, 81679 M\"{u}nchen, Germany}

\altaffiltext{\ExcellenceCluster}{Excellence Cluster Universe, Boltzmannstr.\ 2, 85748 Garching, Germany}

\altaffiltext{\Caltech}{California Institute of Technology, MS 249-17, 1216 E. California Blvd., Pasadena, CA, USA 91125}

\altaffiltext{\JPL}{Jet Propulsion Laboratory, California Institute of Technology, Pasadena, CA 91109, USA}

\altaffiltext{\Berkeley}{Department of Physics, University of California, Berkeley, CA, USA 94720}

\altaffiltext{\UFlorida}{Department of Astronomy, University of Florida, Gainesville, FL 32611}

\altaffiltext{\Colorado}{Department of Astrophysical and Planetary Sciences and Department of Physics,
University of Colorado,
Boulder, CO, USA 80309}

\altaffiltext{\SantaCruz}{SantaCruz UCO/Lick Observatories, University of California, Santa Cruz 95065, USA}

\altaffiltext{\Davis}{Department of Physics, University of California, One Shields Avenue, Davis, CA, USA 95616}

\altaffiltext{\NASA}{Department of Space Science, VP62,NASA Marshall Space Flight Center,Huntsville, AL, USA 35812}

\altaffiltext{\Arizona}{Steward Observatory, University of Arizona, 933 North Cherry Avenue, Tucson, AZ 85721}

\altaffiltext{\MPE}{Max-Planck-Institut f\"{u}r extraterrestrische Physik,Giessenbachstr.\ 85748 Garching, Germany}

\altaffiltext{\CaseWestern}{Physics Department, Center for Education and Research in Cosmology and Astrophysics, 
Case Western Reserve University,Cleveland, OH, USA 44106}

\altaffiltext{\Minnesota}{Department of Physics, University of Minnesota, 116 Church Street S.E. Minneapolis, MN, USA 55455}

\altaffiltext{\STScI}{Space Telescope Science Institute, 3700 San Martin Dr., Baltimore, MD 21218}

\altaffiltext{\ArtInstChicago}{Liberal Arts Department, School of the Art Institute of Chicago, 112 S Michigan Ave, Chicago, IL, USA 60603}

\altaffiltext{\Yale}{Department of Physics, Yale University, P.O. Box 208210, New Haven,CT, USA 06520-8120}

\altaffiltext{\LosAlamos}{T-2, MS B285, Los Alamos National Laboratory, Los Alamos, NM 87545}

\altaffiltext{\KIPAC}{Kavli Institute for Particle Astrophysics and Cosmology 452 Lomita Mall, Stanford University, Stanford, CA, 94305}

\altaffiltext{\Stanford}{Department of Physics, Stanford University, Stanford, CA, 94305}

\altaffiltext{\SLAC}{SLAC National Accelerator Laboratory, 2575 Sand Hill Rd., MS 29, Menlo Park, CA, 94025}

\altaffiltext{\BCCP}{Berkeley Center for Cosmological Physics, Department of Physics, University of California, and Lawrence Berkeley 
National Labs, Berkeley, CA, USA 94720}

\begin{abstract}
We compare cosmic microwave background lensing convergence maps derived from South Pole Telescope (SPT) data with galaxy
survey data from the Blanco Cosmology Survey, the Wide-field Infrared Survey Explorer, and a new large Spitzer/IRAC field designed to overlap
 with the SPT survey. Using optical and infrared catalogs covering between 17 and 68 \degs \  of sky, 
we detect correlation between the SPT convergence maps and each of the galaxy density maps at $>4 \sigma$,
with zero correlation robustly ruled out in all cases. 
The amplitude and shape of the cross-power spectra are in good agreement with theoretical expectations and the measured galaxy bias is 
consistent with previous work.  
The detections reported here utilize a small fraction of the full 2500 \degs \ SPT survey data and serve 
as both a proof of principle of the technique and an illustration of the potential
of this emerging cosmological probe.

\end{abstract}


\section{Introduction}

Gravitational lensing of the primordial anisotropies of the cosmic microwave background (CMB)
imprints information about the density fluctuations 
between $z \sim 1100$ and the present day onto the observed
CMB fluctuations.  This information can be extracted by measuring
the induced correlation between initially independent spatial modes of the 
CMB and used to construct a map of the lensing convergence field. 
This field is closely related to the projected gravitational potential.

The statistics of CMB lensing maps provide 
powerful constraints on cosmological parameters
 \citep{lesgourgues06, deputter09}. Such maps can also be
combined with other tracers of large-scale structure to test 
cosmological models and constrain properties of the tracer
population.  One such analysis involves cross-correlating lensing maps with galaxy catalogs. 
The correlation measures the average lensing signal from the dark matter halos that host the galaxies 
and can be used to determine the halo bias. These measurements test models of the time evolution of 
cosmic density fluctuations and of primordial non-Gaussianity \citep{dalal08,jeong09}.

CMB lensing is a young field.  The first detections---using lensing-galaxy cross-correlations---were reported relatively recently
\citep{smith07, hirata08}. The first detections of lensing in CMB data alone have come from
high-resolution CMB experiments,
both through non-Gaussianity \citep{das11, vanengelen12} and through
smearing of the acoustic peaks \citep{reichardt09a,das11b, keisler11}. Within the next 
year, further advances are expected: the Planck satellite
\citep{planck06} will create all-sky  convergence maps \citep{hanson11}, and the now-completed 2500 \degs \ South
Pole Telescope (SPT) survey \citep{carlstrom11} will produce complementary lensing maps that
have significantly higher
signal-to-noise per mode than the Planck maps. 

Here we present the results of the cross-correlation of several galaxy 
populations with convergence maps from the SPT. In contrast to previous detections of this cross-correlation that utilized a 
large fraction of the sky with low signal-to-noise lensing maps \citep{smith07, hirata08}, 
we use convergence maps that have signal-to-noise greater than 1 on degree scales but 
significantly less sky area; the results presented are from two $\sim100$ \degs \ SPT fields.
A significant cross-correlation is detected
between the  convergence maps and maps of galaxy density constructed
from optical and infrared (IR) catalogs in each of the two fields.

The paper is structured as follows.
We first discuss the underlying theory and provide a brief overview of the process of making convergence maps from CMB data. 
We next describe the catalogs, real and simulated, that we cross-correlate with 
the lensing maps. We conclude
with a discussion of the results and the potential of upcoming
large CMB lensing data sets and large-area galaxy surveys.

\section{Theory}

The CMB lensing convergence in a direction $\nhat$ on the sky is given in terms of the matter fluctuations by a line-of-sight integration, 
\begin{equation}
\kappa(\nhat) = \int d\chi \, W^\kappa(\chi) \delta( \chi\nhat,z(\chi)),
\end{equation}
where  $\delta(\mathbf{r}, z)$ is the fractional matter over-density at comoving position $\mathbf{r}$ and redshift $z$, and  the distance kernel is \citep{cooray00b, song03}
\begin{equation}
W^\kappa(\chi) = {3 \over 2} \Omega_m H_0^2 {\chi \over a(\chi)}  {\chicmb - \chi \over \chicmb }. 
\end{equation}
Here, $\Omega_m$ is the matter density relative to the critical density evaluated today, $H_0$ is the Hubble parameter today, $a(\chi)$ is the cosmological scale factor,  $\chicmb \simeq 14\,$Gpc is the comoving distance to the CMB recombination surface, and we have assumed a spatially flat Universe.

Under the assumption that observed galaxies are biased tracers of mass fluctuations, the observed fractional galaxy over-density in a direction $\nhat$ is given by
\begin{equation}
g({\nhat}) = \int d\chi\, W^g(\chi) \delta(\chi\nhat , z(\chi)),
\end{equation}
where the distance kernel is
\begin{equation}
W^g(\chi) = {1 \over{ \left[ \int dz^\prime \, {dN(z^\prime) \over dz^\prime} \right] } } {dz\over d\chi} {dN(z) \over dz}   b(\chi). 
\end{equation}
Here, $dN(z)/dz$ is the distribution of galaxies in redshift and $b(\chi)$ is the bias of the galaxies relative to the dark matter density, assumed here to be independent of scale.

The cross power between the convergence and the galaxy over-density at a multipole $L$ is given in the Limber approximation \citep{limber53, kaiser92} by 
\begin{equation}
C_L^{\kappa g} = \int dz\, {d\chi \over dz} {1\over\chi^2} W^\kappa(\chi) W^g(\chi) P\left(k = {L \over \chi}, z\right),
\label{eq:clcross}
\end{equation}
where the matter power spectrum today, $P(k,0)$ is given by $\langle \delta^\star(\kvec) \delta(\kvec^\prime) \rangle = (2\pi)^3 \delta_{\mbox{\tiny Dirac}}(\kvec - \kvec^\prime) P(k,0)$ under the assumption of independent Fourier modes $\kvec$ and $\kvec^\prime$.  The amount of cross-correlation is thus determined by the overlap between the two kernels, weighted by the matter power spectrum.  The factors which determine this overlap, namely ${1 \over \chi} W^{[g, \kappa]}(z) (P(k, z) d\chi / dz)^{1/2}$, are plotted for the CMB lensing convergence and several galaxy populations in the bottom panel of Figure  \ref{fig:dndz}.

\section{Convergence Maps from CMB Lensing}

CMB maps of two fields, one centered at (RA,DEC) = (23h30m,-55d) 
and the other at (5h30m,-55d), were constructed using
150 GHz data from the SPT survey. These fields together encompass
185 \degs \  and are the deepest SPT fields to date 
as they were observed with roughly twice the time per unit area as the
rest of the SPT survey. 
In addition to the 2008 data used in 
\citet{vanengelen12} (hereafter V12), the maps used in this analysis include data from 
the 2010-2011 observing seasons. As such, the resulting lensing maps
are $\sim 10\%$ lower in noise  
than the typical maps to be expected from the rest of the survey.

Convergence maps were constructed using a quadratic 
estimator \citep{hu02a}, as outlined in V12.
Point sources with signal-to-noise greater than 6 
were masked, while clusters detected via the thermal Sunyaev-Zeldovich (tSZ) effect were masked if they had a signal-to-noise greater than 6 in the 2008 data alone. 
This cross-correlation study is less sensitive to foreground contamination and 
offsets in power than the V12 analysis; therefore, the 
temperature maps were filtered to include modes from $\ell=$ 1200--4000, a 
larger range than
the $\ell=$ 1200--3000 used in V12.

The effect of foreground contamination is expected to be small for this
analysis. Residual foreground emission in the CMB map that is correlated with the 
galaxy distribution is likely to lead
to negative cross-correlations on large scales: excess small-scale power
fluctuations in the temperature map are identified by the
quadratic estimator as large-scale under-dense regions which
`squeeze' the sky. This produces an anti-correlation of true large 
scale structure with residual foreground contamination in the lensing reconstruction. 
Conversely, dust in our Galaxy could also be interpreted by the lensing 
estimator as large-scale under-dense regions 
while also
suppressing the galaxy counts, leading to a net positive correlation. 
We discuss tests for foreground contamination in Section 6.

The measured CMB lensing signal was calibrated 
with simulations. As outlined in V12, simulated lensed CMB skies including contributions 
from faint IR sources and tSZ from low-mass
galaxy clusters (both modeled as Gaussian random fields)
were added to noise generated from SPT timestreams. 
The simulations were filtered and analyzed in the same way as the
real data. We normalize the reconstructed convergence maps as a function of $L$ 
by comparing the cross-spectra of the input and output maps to the power spectra of the input
maps. The simulated lensing maps are also used to estimate the uncertainty in the
cross-correlation.

\section{Optical and IR-Selected Catalogs}

We use optical- and IR-selected catalogs to create maps of the fractional galaxy over-density, $(N-\bar{N})/\bar{N}$, 
where $N$ is the galaxy counts in a 1$'$ cell. A bright magnitude threshold is applied to each catalog to 
reduce 
contamination from Galactic objects, while a faint magnitude limit is set to ensure uniformity in source
detection. We exclude regions around bright stars found in the 2MASS source catalog \citep{skrutskie06}. 

Optical catalogs were produced using data from the Blanco Cosmology Survey (BCS, \citep{desai12}).
The survey consists of two fields, a 5h and a 23h field, both contained within the SPT footprint. 
For this analysis, we extracted sources from images 
reduced using the pipeline described in \citet{rest05a} and 
\citet{miknaitis07}. Source detection was performed using SExtractor v2.8.6 \citep{bertin96},
 and the photometry was calibrated using Stellar Locus Regression \citep{high09}.
 We restrict the analysis to areas that are complete to at least
 $i_{AB}=22.5$ and construct catalogs from all sources $19.5<i_{AB}<22.5$.

\begin{figure}
\epsscale{1.0}
\plotone{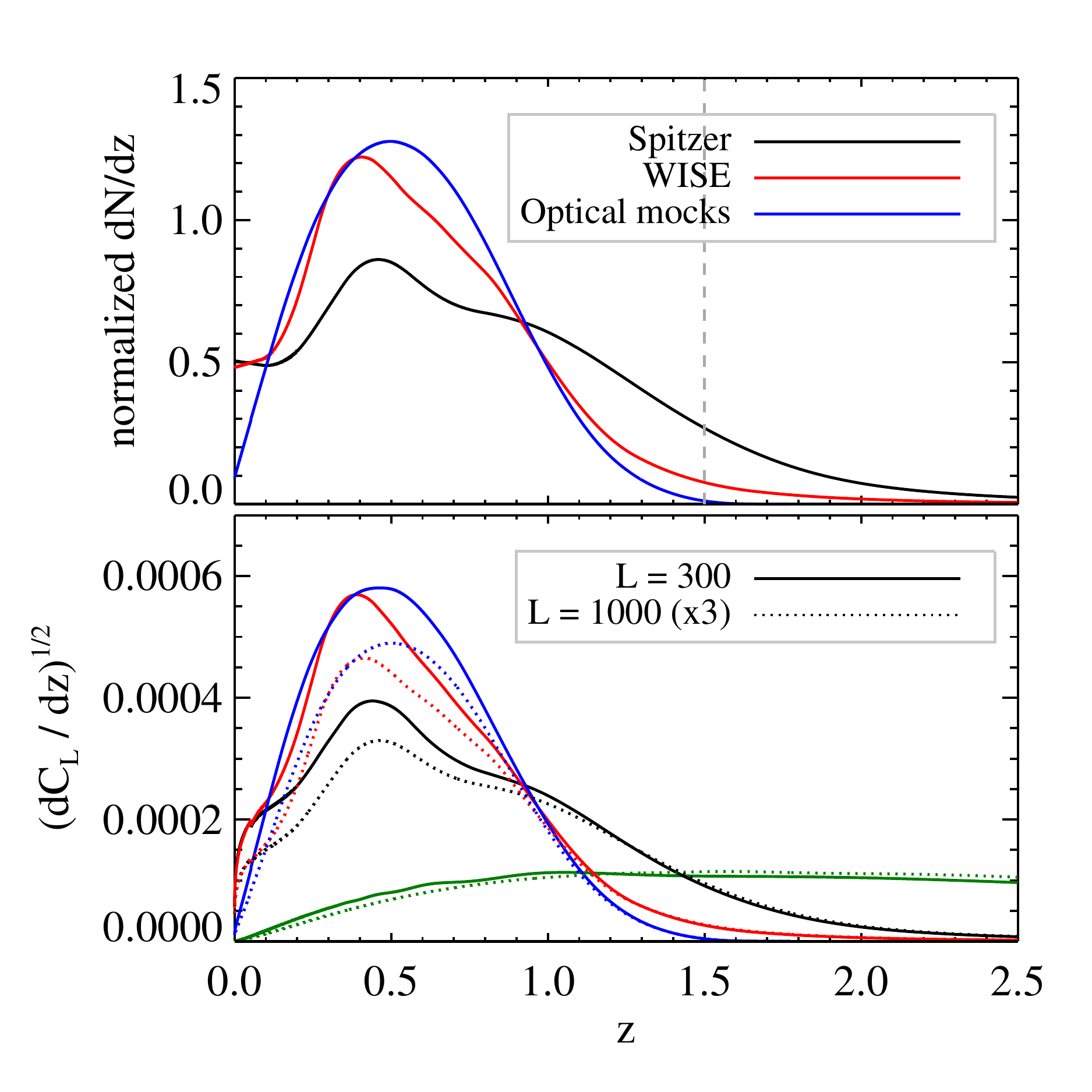}
\caption{Redshift distribution for the galaxy samples are shown in the top panel. 
Mock catalogs (Section 5) are used to estimate the distribution for the optically-selected galaxies, 
while IR-selected galaxy distributions are based on photometric redshift estimates 
from \citet{brodwin06}. The mock catalogs extend to $z = 1.3$, a sufficient redshift limit for the 
optical catalogs under consideration. The vertical dashed line at $z = 1.5$ indicates 
the maximum redshift of the \citet{brodwin06} spectroscopic training sample. 
The quantities shown in the bottom panel, which includes the CMB lensing power (green lines), 
are the curves that one would multiply to obtain  $dC_L^{\kappa g}/dz$, the integrand of equation 5, 
at $L = 300$ (solid curves) and $L = 1000$ (dotted curves), assuming galaxy bias $b(z) = 1$. 
The dotted curves have been multiplied by 3 for clarity.
}
\label{fig:dndz}
\end{figure}
\nocite{brodwin06}

In the 23h field, we also use data from the SPT Spitzer Deep Field (Ashby, private communication), 
an ongoing survey of 100 \degs \  at 3.6 and 4.5$\mu$m. Data reduction followed the methods of \citet{ashby09}, except, 
owing to the new survey's large size, photometry from individual 2 \degs \ sub-fields was merged in catalog space instead of extracted from a single monolithic mosaic. 
Here we analyze 32 \degs \ of existing coverage and include all sources with 
4.5$\mu$m magnitude between 15 and 17 Vega.  

As a second catalog in the 5h field we use the sources from  
the Wide-field Infrared Survey Explorer (WISE) \citep{wright10} all-sky data release. 
We restrict this analysis to areas with W1 band exposure times between 231-616s, 
exclude regions with potentially corrupted objects as identified by the WISE  
pipeline, and include all sources with W1 magnitude between 15 and 17 Vega.
 Although the WISE selection function is complicated at these magnitudes 
(owing to source confusion\footnote[1]{http://wise2.ipac.caltech.edu/docs/release/allsky/expsup/sec6\_5.html} 
and spatial variability in the imaging depth), 
we find it necessary to use these faint sources to obtain sufficient numbers of high-redshift galaxies. 
Despite its limitations, the WISE sample is a particularly interesting test case, 
as WISE is an all-sky survey, and this analysis will soon be possible over the entire SPT survey area. 

\begin{figure*}
\epsscale{1.0}

\plotone{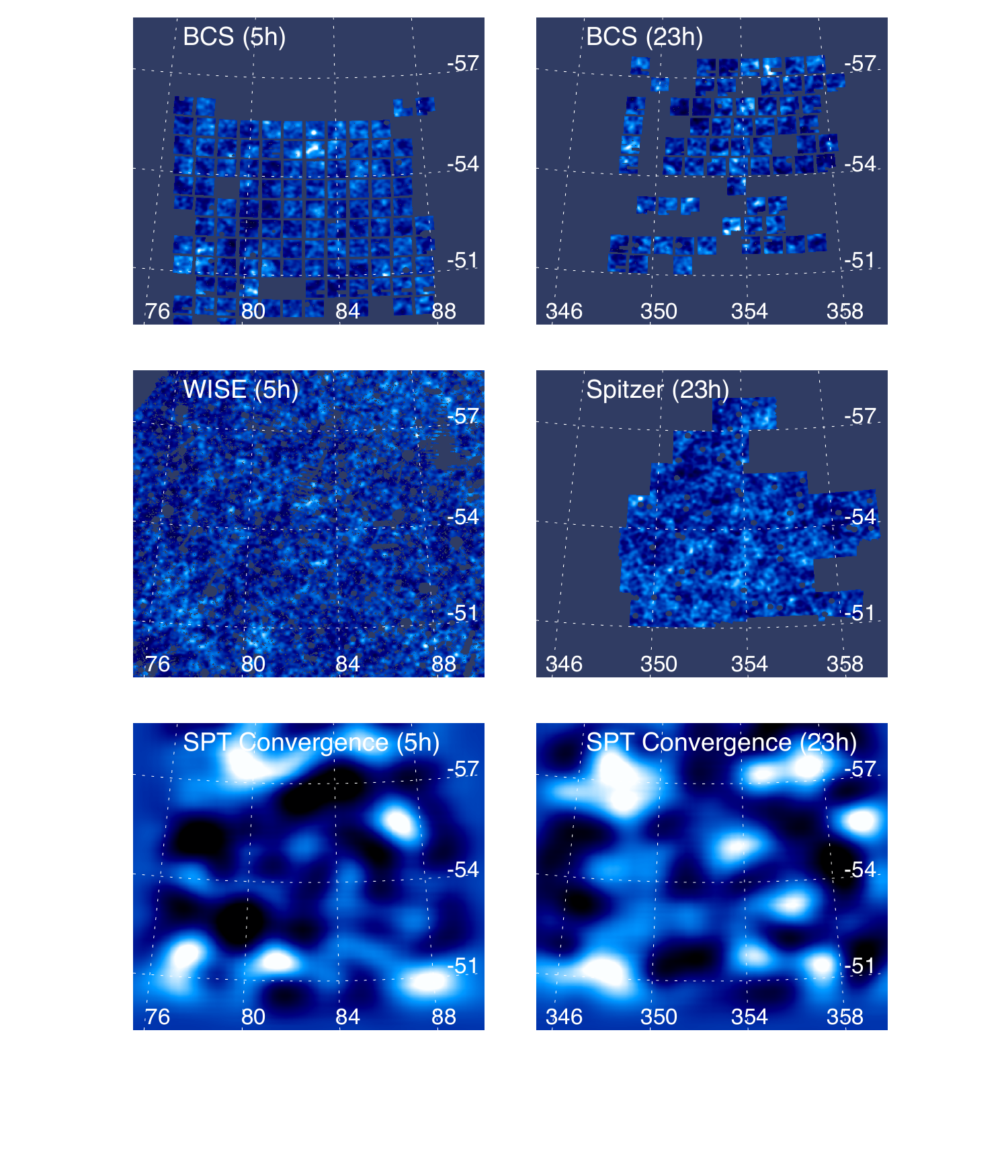}

\caption{ The top four panels show fractional galaxy over-density maps smoothed to 6$'$ resolution. 
The bottom two panels show the lensing maps smoothed to 1$^{\circ}$ resolution. 
At this scale, the signal-to-noise in the lensing maps is slightly greater than 1. While the lensing maps shown here 
have been smoothed to highlight real mass fluctuations, it is clear from the cross-spectra
there is substantial information at smaller scales as well.  The color scale is $\pm$  80\% of 
the maximum deviation from the mean in each map.}
\label{fig:maps}
\end{figure*}

The expected redshift distributions for the galaxy catalogs are shown in Figure \ref{fig:dndz}. 
Summary statistics are presented in Table 1, and galaxy density maps are shown in Figure \ref{fig:maps}.

\section{Mock Catalogs}

Mock catalogs were used to estimate both the expected level of cross-correlation between CMB convergence maps and
 galaxy catalogs and the potential contamination from tSZ. The simulated galaxy sample is drawn from a 220 \degs \
 lightcone populated with galaxies in the redshift range 0$-$1.3.  The underlying dark matter distribution is based on a cosmological 
simulation of 1 $h^{-1}$Gpc; this simulation is a single `Carmen' 
simulation from the Large Suite of Dark Matter Simulations project 
(LASDAMAS) (see http://lss.phy.vanderbilt.edu/lasdamas/).
The Adding Density Determined GAlaxies to Lightcone Simulations (ADDGALS) algorithm 
is run to assign galaxies to the dark matter particles in a way that reproduces the known luminosities, colors, and 
clustering properties of galaxies.  The simulated catalogs have previously been used for tests of cluster finding 
\citep{koester07a}, photometric redshifts \citep{gerdes10}, and spectroscopic followup \citep{cunha11}.
A full description of the algorithm and the simulated sky catalog will be presented in Wechsler et al. and Busha et al. (in preparation).
The simulated CMB lensing map based on the same underlying dark matter distribution was produced using multiple-plane ray tracing 
methods similar to those of \citet{hilbert09}, with modifications to produce maps with the proper lensing kernel at the 
redshift of the CMB. Thermal SZ was included in the simulation as in \citet{biesiadzinski12}: tSZ profiles were
generated at the location of the halos using the \citet{arnaud10} model and projected
along the line-of-sight to create a simulated tSZ map.

Owing to the finite size of the simulation box, 
the mock catalogs and the CMB lensing map only extend to $z
\sim 1.3$. For optical catalogs, this is not a serious
limitation  but there could be substantial cross-correlation coming from
higher redshifts in the IR-selected catalogs. Also, these mock catalogs
do not include AGN (which could enhance the cross-correlation), stars
(which would not be correlated with a CMB lensing map), or the effects of source confusion (which would depress the cross-correlation).

The simulated galaxy catalogs were constructed using the same selection criteria
used for the real optical catalog. Maps of the fractional galaxy over-density were 
created and cross-correlated with the associated lensing convergence map.

\begin{table*}[t]
\caption{Field parameters and correlation statistics}

\resizebox{18cm}{!}{

\begin{tabular}{c c c c c c c c}
Field & Area ($\mathrm{deg}^{2}$) & Density ($\frac{\mathrm{sources}}{\mathrm{deg}^{2}}$) & $A$ ($C_{L}\times10^{-7}$)  &  $n$ & $\chi^2$ (Best fit) &  $\Delta \chi^2 (0)$ & Bias \\
\hline
WISE (5h)  & 68.1 & 6.9$\times 10^{3}$ & $0.19 \pm 0.05 $ &  $-1.2 \pm 0.3$ & 8.8 & 19.6 & $0.9 \pm 0.2$ \\  
BCS (5h) & 27.0 & 2.5$\times 10^{4}$ & $0.27 \pm 0.06$ &$-1.8 \pm 0.3$ & 11.3 & 23.5 & $1.2 \pm 0.3$ \\  
BCS (23h) & 16.9 & 2.35$\times 10^{4}$ & $0.24 \pm 0.07$ & $-1.7 \pm 0.3$ & 9.6 & 17.5& $1.1 \pm 0.3$ \\
Spitzer (23h) & 29.8 & 1.4$\times 10^{4}$ & $0.33 \pm 0.07$ & $-1.6 \pm 0.2$ & 13.7 & 28.9 & $1.7 \pm 0.3$\\ 
\hline
\end{tabular}
}

\begin{tablenotes}[para]
Galaxy catalog properties and results of
power-law fits to lensing-galaxy 
cross spectra.  For each catalog, we report best-fit
amplitude and power law index, $\chi^2$ of the best fit, the difference in $\chi^2$ from best-fit to a model with zero cross-correlation, and the bias. Note: we report the 
weighted area of the galaxy density maps multiplied by the lensing apodization mask.

\end{tablenotes}
\end{table*}

\section{Results and Discussion}
\begin{figure}[!]
\epsscale{1.0}
\plotone{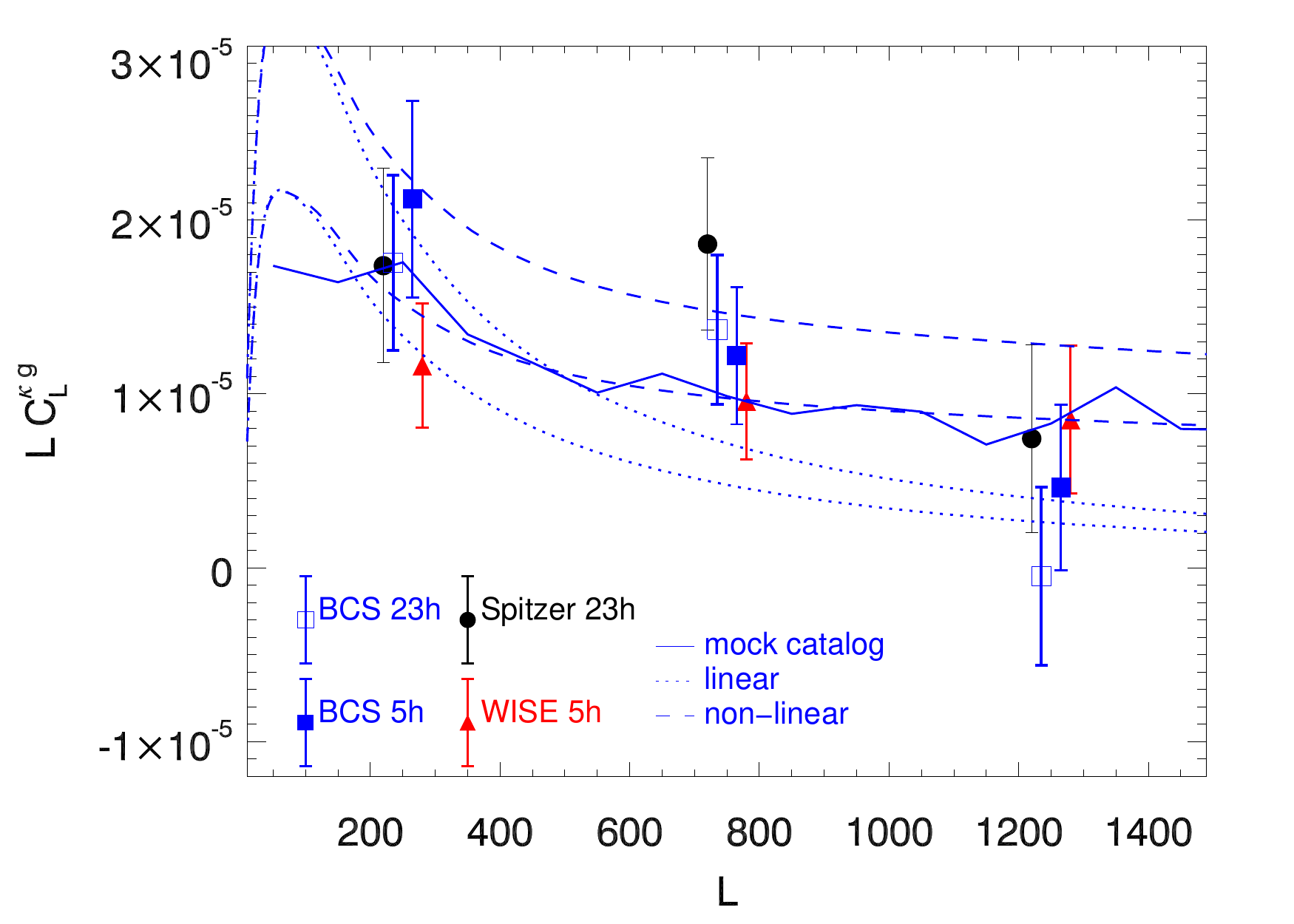}
\caption{Cross-spectra of galaxy number density maps with SPT lensing convergence maps.
Spectra are calculated in 14 bins from 
$L=150-1450$, but are shown combined into 3 bins for display
purposes. The solid blue line is calculated using the mock catalogs and convergence maps. 
The dashed and dotted lines are
obtained using the Limber approximation and the redshift distribution
from the mock catalogs for the non-linear and linear power spectrum,
respectively. The upper and lower curves for each model correspond to a redshift-independent galaxy bias of 
1.0 and 1.5, respectively. For clarity we only plot models for the optical catalogs. 
Predictions for the IR samples are within $15\%$ of the plotted curves.
}
\label{fig:cross_spectra}
\end{figure}

We calculate cross-power spectra between the CMB lensing maps and the maps of 
fractional galaxy over-density from each of the galaxy catalogs. 
Given the high source density in the galaxy catalogs and the cuts designed 
to enhance uniformity in source selection, we expect the noise in these 
correlations to be dominated by the lensing reconstruction. Noise estimates 
are obtained by cross-correlating the galaxy maps with 50 simulated lensing 
maps (including realistic noise). Results are shown in Figure \ref{fig:cross_spectra}.

Significant correlated signal\footnote[2]{Signal-to-noise is defined as $\sqrt{\Delta \chi^2(0)}$. $\Delta \chi^2(0)$ 
is the difference in $\chi^2$ from best-fit to a model with zero cross-correlation} 
($\sim 4.2-5.3\sigma$) is detected for every catalog, 
and zero cross-correlation is ruled out robustly in all cases. 
Additionally, we note a conservative magnitude cut is used in this Spitzer 
analysis---extending the catalogs one magnitude fainter increases the correlation to 
$>8.8 \sigma$---but, as the properties of this deeper catalog are still being characterized, 
we leave detailed study of this sample to future work.

We test for contamination from Galactic cirrus by cross-correlating lensing maps and galaxy maps with predictions for 
the Galactic dust emission
\citep{finkbeiner99} and find no correlation. 
We also place a limit on the systematic bias owing to tSZ using the simulated tSZ map 
described in Section 5. We mask an identical density of clusters as in the real CMB maps 
and process this map through
the quadratic estimator used to create the convergence maps. 
The resulting maps of tSZ leakage are 
cross-correlated with the mock galaxy density maps. 
The tSZ leakage has a small effect, suppressing the correlation by $\sim5\%$.

\begin{figure}
\epsscale{1.0}
\plotone{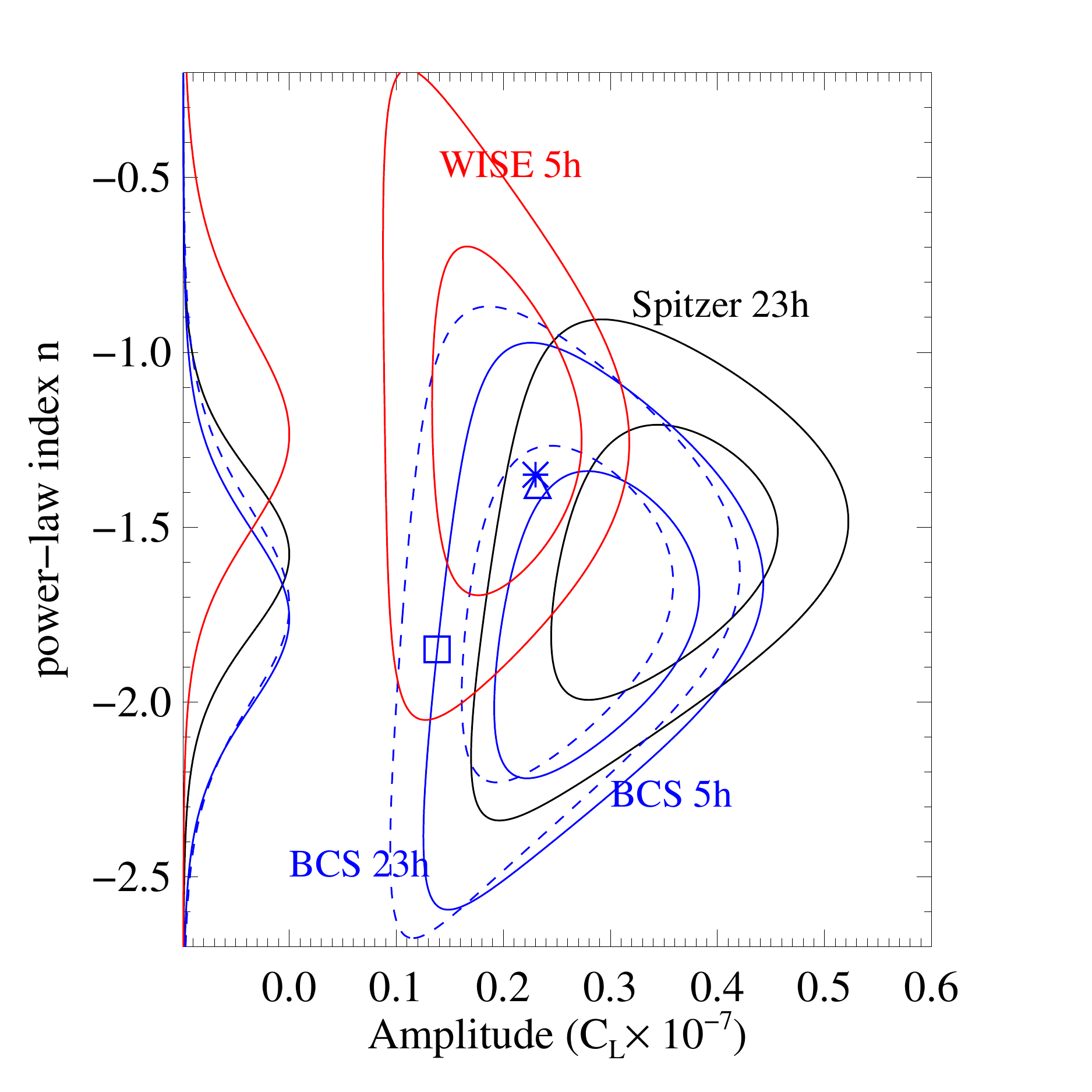}
\caption{Power-law fits
to SPT/galaxy density cross-spectra. 
The marginalized 1-d probability distributions for the power-law
indices are shown projected onto the vertical axis. 
Also shown are the predictions from mock catalogs for 
optically selected galaxies (star), 
and predictions using the mock redshift distribution 
and using non-linear (triangle) and linear (square) power spectra, 
assuming a bias of 1. }
\label{fig:chi_2d}
\end{figure}

A power-law is fit to each measured cross spectrum: $C_L =
A (L/500)^{n}$.  In all cases a simple 
power law is a good fit and the best-fit amplitudes are significantly above zero.
The data are consistent with a single power law describing all data sets.
The constraints on amplitude and power-law index for each 
cross-spectrum are shown in Figure \ref{fig:chi_2d} and summarized in Table 1. 
In Figures \ref{fig:cross_spectra} and \ref{fig:chi_2d} we also plot the 
results from the mock catalogs and convergence maps. Additionally we plot the 
theoretical expectation (calculated using the CAMB software package, \citealt{lewis00}) for both the 
non-linear and linear matter power spectrum using the galaxy redshift distributions from the mock catalogs.
We also constrain the redshift-independent bias assuming non-linear evolution, by fixing the power law index to the non-linear value and evaluating the 1D amplitude probability distribution relative to the $b=1$ value.
These results are consistent with previous measurements \citep{delatorre07}.

The measured power-law slopes are in rough agreement with expectations for large-scale 
structure that is mainly in the non-linear regime. The angular
correlation function of optical galaxies is reasonably well-fit
by a simple power law with $w(\theta) \propto \theta^{-0.77}$
(e.g., Peebles 1975\nocite{peebles75}), which corresponds to $n=-1.23$.
The fluctuations in the cosmic IR background are also
well-fit by a power law with $n=-1.2$ \citep{addison12,planck11-6.6_arxiv,reichardt11}.
It is possible that at low $L$ the cross-correlation is probing scales that are 
still in the linear regime, while at high-$L$ the clustering power spectrum which dominates has a steeper slope, 
resulting in a good approximation of a power law.  The data are not yet sufficiently 
constraining to test this hypothesis. 

Modeling the signal further is complicated by 
the lack of precise redshift information for each catalog, residual
contamination of the galaxy catalogs by Galactic objects, and incompleteness due to source confusion. 
These are not 
fundamental limitations, and work is ongoing to both improve and better characterize these galaxy catalogs.
Even in the presence of these limitations, the cross-spectra generally appear to agree 
with the expectations based on both the mock catalogs and theory.

\section{Conclusions}

 We have demonstrated that CMB lensing convergence maps are strongly
correlated with other tracers of large scale structure. 
These results are derived from a small fraction ($<$ 10\%) of
the now completed 2500 \degs \ SPT survey, and work is underway to create convergence maps of the full survey. 
Realizing the full potential of the cross-correlation analyses will require deep galaxy surveys over 
the complete survey area. The positive correlation with WISE is promising, 
as all-sky WISE catalogs have recently become available. The SPT-Spitzer Deep Field will ultimately cover three times the area 
presented here.  This same field is scheduled for observation with 
{\em Herschel}-SPIRE
and will soon be imaged with SPTpol \citep{bleem12b}, resulting in even deeper lensing maps.
In the next few years, the Dark Energy Survey and VISTA Hemisphere Survey are expected to 
fully cover the SPT survey at optical and near-IR wavelengths.

With existing and planned high signal-to-noise, large-area CMB lensing maps, 
the near future promises
rapid evolution in the state of CMB lensing analyses. 
Combining the lensing maps with multi-wavelength information from upcoming
galaxy surveys will allow exploration of galaxy formation models out to 
high redshift and of structure growth through measurement of the amplitude and 
evolution of cosmic density fluctuations.
Careful modeling of the survey galaxy populations, 
sources of bias in the lensing reconstructions, and 
theoretical expectations for the signal will allow 
these cross-correlation analyses to achieve their full potential.

\begin{acknowledgements}
The SPT is supported by 
grants ANT-0638937 and ANT-0130612, 
with partial support provided by 
PHY-0114422, the Kavli Foundation, and the 
Moore Foundation. Work at McGill is supported by
NSERC, the CRC program, and CIfAR, 
and at Harvard by grant AST-1009012.
R.\ Keisler acknowledges NASA Hubble Fellowship grant HF-51275.01, 
B.A.\ Benson a KICP Fellowship,
M.\ Dobbs an Alfred P. Sloan Research Fellowship,
L.\ Shaw grant AST-1009811,
R.\ Wechsler DOE contract DE-AC02-76SF00515, 
A.\ Vallinotto DOE contract DE-AC52-06NA25396(LA-UR-12-20137), 
and O.\ Zahn a BCCP fellowship.
This publication uses data from the Wide-field Infrared Survey Explorer, a joint project of UCLA, and JPL/Caltech, funded by NASA and uses data provided by NOAO PI 2005B-0043, distributed by the NOAO Science Archive. 
NOAO is operated by AURA under cooperative agreement with the NSF. 
This work is based in part on observations made with the Spitzer Space Telescope, operated by JPL, Caltech under a contract with NASA. 

\end{acknowledgements}

{\it Facilities:}
\facility{Blanco (MOSAIC)},
\facility{Spitzer (IRAC)},
\facility{South Pole Telescope},
\facility{Wide-field Infrared Survey Explorer}

\bibliography{../../BIBTEX/spt}

\end{document}